\documentclass[prb,preprint,aps,superscriptaddress,longbibliography]{revtex4-1}
\usepackage[]{color}
\usepackage{appendix}
\usepackage{graphicx, epstopdf}
\usepackage{subfigure}
\usepackage{float}
\usepackage{multirow}
\usepackage{amsmath}
\usepackage{amsfonts}
\usepackage{amssymb}
\usepackage{bm}
\usepackage{subfigure}
\usepackage{caption}
\usepackage{hyperref}
\usepackage{xr}
\usepackage{lineno}
\begin{document}
\title{CTGNN: Crystal Transformer Graph Neural Network for Crystal Material Property Prediction}


\author{Zijian Du}
\thanks{These two authors contributed equally to this work.}
\affiliation{School of Information Science and Technology, Fudan University, Shanghai 200433, China.}
\affiliation{Department of Physics, Fudan University, Shanghai 200433, China.}

\author{Luozhijie Jin}
\thanks{These two authors contributed equally to this work.}
\affiliation{School of Information Science and Technology, Fudan University, Shanghai 200433, China.}

\author{Le Shu}
\affiliation{School of Information Science and Technology, Fudan University, Shanghai 200433, China.}

\author{Yan Cen}
\email{cenyan@fudan.edu.cn}
\affiliation{Department of Physics, Fudan University, Shanghai 200433, China.}

\author{Yuanfeng Xu}
\affiliation{School of Science, Shandong Jianzhu University, Jinan 250101, Shandong, China}

\author{Yongfeng Mei}
\affiliation{Department of Materials, Fudan University, Shanghai 200433, China.}

\author{Hao Zhang}
\email{zhangh@fudan.edu.cn}
\affiliation{School of Information Science and Technology, Fudan University, Shanghai 200433, China.}
\affiliation{Department of Optical Science and Engineering and Key Laboratory of Micro and Nano Photonic Structures (Ministry of Education), Fudan University, Shanghai 200433, China.}
\affiliation{State Key Laboratory of Photovoltaic Science and Technology, Fudan University, Shanghai 200433, China}

\date{\today}

\begin{abstract}
The combination of deep learning algorithm and materials science has made significant progress in predicting novel materials and understanding various behaviours of
materials. Here, we introduced a new model called as the Crystal Transformer Graph Neural Network (CTGNN), which combines the advantages of Transformer model and graph neural networks to address the complexity of structure-properties relation of material data. Compared to the state-of-the-art models, CTGNN incorporates the graph network structure for capturing local atomic interactions and the dual-Transformer structures to model intra-crystal and inter-atomic relationships comprehensively. The benchmark carried on by the proposed CTGNN indicates that CTGNN significantly outperforms existing models like CGCNN and MEGNET in the prediction of formation energy and bandgap properties. Our work highlights the potential of CTGNN to enhance the performance of properties prediction and accelerates the discovery of new materials, particularly for perovskite materials.
\end{abstract}

\flushbottom
\maketitle

\section{Introduction}

Deep learning (DL) and machine learning (ML) has brought significant impacts to a variety of scientific fields such as biology\cite{zhou2020graph}, chemistry\cite{jumper2021highly},physics\cite{kirkpatrick2021pushing} and mathematics\cite{davies2021advancing}. While in materials science, the use of deep learning has led to important progress in material properties prediction\cite{xie2018crystal}, materials generation\cite{xie2021crystal} , etc\cite{PredictingBandGaps,Butler2018,lecun2015deep,wu2020comprehensive}. Several DL models have been developed to capture material modality and predict their properties, such as Crystal Graph Convolutional Neural Network (CGCNN)\cite{xie2018crystal}, MatErials Graph Network (MEGNET)\cite{chen2019graph}, Atomistic Line Graph Neural 
Network (ALIGNN)\cite{choudhary2021atomistic}, improved Crystal Graph Convolutional Neural Networks (iCGCNN)\cite{park2020developing},
OrbNet\cite{qiao2020orbnet}, and similar variants\cite{gasteiger2020directional,gasteiger2020fast,shui2020heterogeneous,schutt2017quantum,anderson2019cormorant,zhang2020molecular,schütt2018schnetpack,jha2018elemnet,westermayr2020combining,wen2021bondnet,isayev2017universal}. They have achieved great success in applications, such as learning properties
from multi-fidelity data\cite{chen2021learning}, discovering stable lead-free hybrid organic–inorganic perovskites\cite{lu2018accelerated}, mapping the
crystal-structure phase\cite{chen2021automating}, and designing material microstructures\cite{lee2021fast}.

Despite the graph-based DL model, the Transformer\cite{vaswani2017attention} model provides a new way to capture the material information, and some models based on Transformers to predict material properties have been developed, such as MatFormer\cite{yan2022periodic}, Graphformer\cite{min2022transformer} , etc. These models integrate the Transformer as the core network, utilizing the connections within graphs as the queries, keys, and values (QKV) in the attention mechanisms, distinguishing them from traditional graph neural networks. Therefore, they lose the conventional graph structure\cite{min2022transformer}. Some other models based on Transformer models use the structures of graph neural networks such as ADA-GNN\cite{Huang2024ada-gnn:}, TG-GNN\cite{PPR:PPR527806}, GATGNN\cite{Louis2020GlobalAB}, etc. But these models further introduce high complexity on the basis of Transformer architecture, which are not conducive to model training. To address the aforementioned limitations, in this work, the Crystal Transformer Graph Neural Network (CTGNN) is proposed, which combines the Transformer structures' message capturing capabilities and traditional inductive bias of GNNs. 

Generally, the GNN-based models extract structural data such as bond length, angles, and neighbour atoms, which are important information to predict the materials properties. In contrast to traditional GNNs which only capture bond length, our proposed CTGNN employs an angular encoder kernel to encode angle features, and the dual-Transformer structures are built, which include one Transformer architecture focusing on intra-crystal interactions to model the immediate chemical environment of atoms, and another to analyze inter-atomic relationships within an atom's neighborhood facilitates a thorough understanding of material behaviors on both local and broader scales. In this work, we conducted a series of ablation experiments to verify the importance of our angular encoding and Transformer architecture in improving model accuracy. We also tested the performance of CTGNN on some widely-used materials database, achieving better results than other models we used for comparison.

\section{Model}
\subsection{Transformer Model}
The Transformer model\cite{vaswani2017attention}, a key component in the CTGNN architecture, is based on the multihead-self-attention mechanisms. The multi-head  design allows the model to efficiently process sequential data in parallel, enhancing its abilities, while the self-attention method allows it to learn relationships between sequences, which is the core of the algorithm.

The self-attention mechanism can be described by  
\begin{equation}
    \text{Attention}(Q, K, V) = \text{softmax}\left(\frac{QK^T}{\sqrt{d_k}}\right)V
\end{equation}
where \(Q\), \(K\), and \(V\) represent queries, keys, and values, while \(d_k\) is the dimension of the key vectors, serving as a reweighting factor that stablizes training process. The multi-head-self-attention allows the model to simultaneously process information from different groups,
\begin{equation}
    \text{MultiHead}(Q, K, V) = \text{Concat}(\text{head}_1, \ldots, \text{head}_h)W^O
\end{equation}
where \(\text{head}_i = \text{Attention}(QW_i^Q, KW_i^K, VW_i^V)\), and \(W^O\) is a parameter matrix.

Then the feed-forward Networks, consisting of two linear transformations with a ReLU activation in between,  are given by
\begin{equation}
    \text{FFN}(x) = \max(0, xW_1 + b_1)W_2 + b_2
\end{equation}
where $W_1$ and $W_2$ are weight matrices, and $b_1$ and $b_2$ are bias vectors. These networks are applied independently to each position in the sequence.

\subsection{CTGNN Model}

\begin{figure}[H]
    \centering
    \includegraphics[width=1\linewidth]{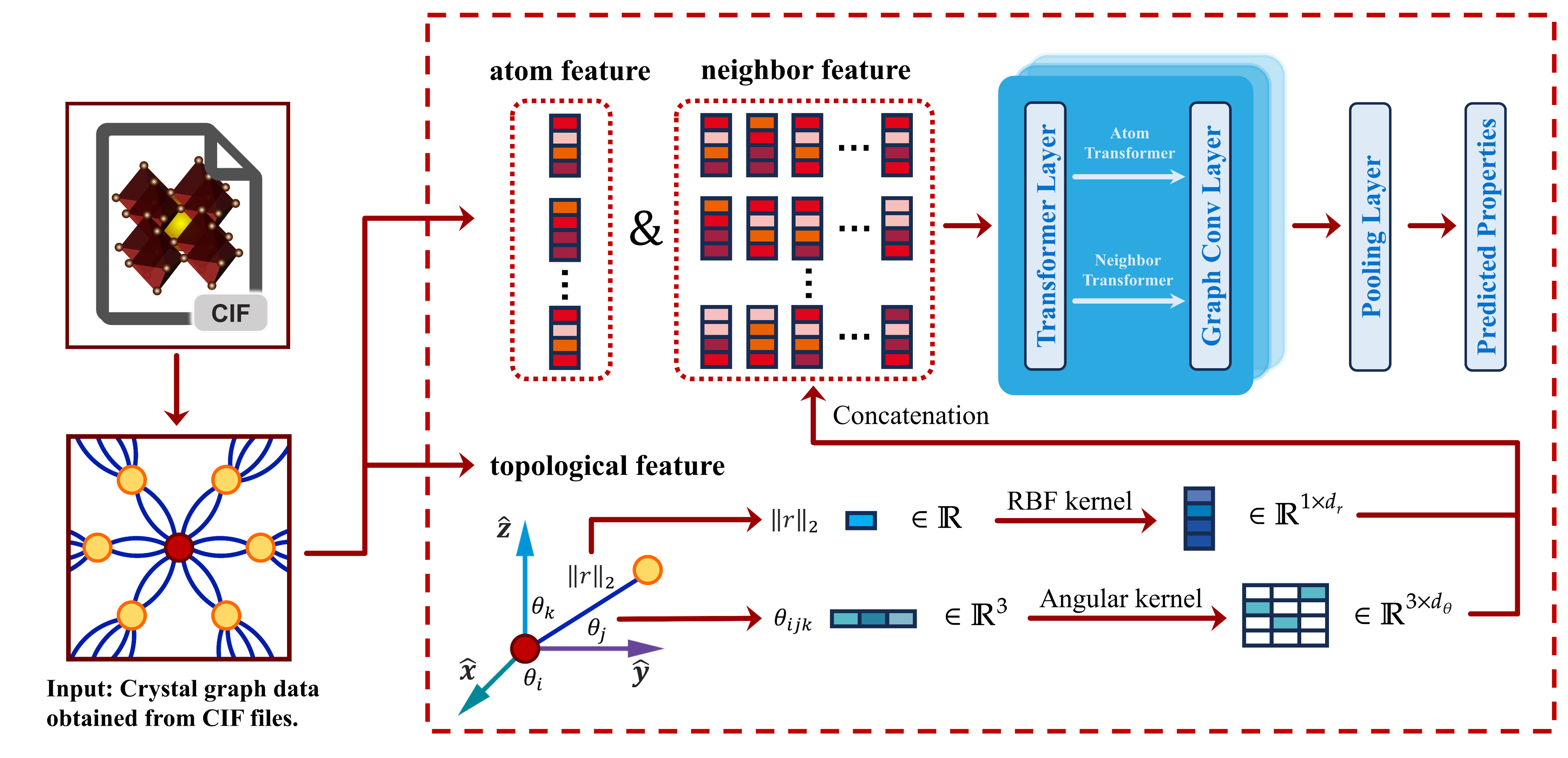}
    \caption{Our CTGNN framework. Transformer Layer denotes as Transformer encoder for atom features and neighbor features. After the Transformer Layers, CGCNN convolution layers are used. And finally pooling and predicting layers are used to get the prediction. Topological features include angular and distances (RBF) information.}
    \label{fig:pero}
\end{figure}

The framework of the proposed CTGNN model is shown in Fig\ref{fig:pero}. In this framework, each atom is denoted as a node while each atom connection as an edge. The node \(i\) in the graph \(G\) is characterized by a vector $v_i$, while the connection between two nodes \(i\),\(j\) is denoted as an edge vector $u(i, j)_{k}$. The node and edge features are updated by Transformer-based graph convolution calculations.

First, the node features are updated by the intra-crystal Transformer layer as,
\begin{equation}
v_i^{(t)} = \text{MultiHead}(Q, K, V) + v_i^{(t-1)}
\end{equation}
\begin{equation}
Q = K = V = \text{Linear}(v_i^{(t-1)})
\end{equation}
where \(\text{MultiHead}(Q, K, V)\) is the multi-head self-attention mechanism defined as:
\begin{equation}
\text{Attention}(Q, K, V) = \text{softmax}\left(\frac{QK^T}{\sqrt{d_k}}\right)V
\end{equation}
\begin{equation}
\text{MultiHead}(Q, K, V) = \text{Concat}(\text{head}_1, \ldots, \text{head}_h)W^O
\end{equation}
\begin{equation}
\text{head}_i = \text{Attention}(QW_i^Q, KW_i^K, VW_i^V)
\end{equation}
After the multi-head attention, the feed-forward network (FFN) is applied,
\begin{equation}
v_i^{(t)} = \text{FFN}(v_i^{(t)})
\end{equation}
\begin{equation}
\text{FFN}(x) = \max(0, xW_1 + b_1)W_2 + b_2
\end{equation}

Similarly, the edge features are updated by the inter-atomic Transformer layer,
\begin{equation}
u(i,j)_k^{(t)} = \text{MultiHead}(Q, K, V) + u(i,j)_k^{(t-1)}
\end{equation}
\begin{equation}
Q = K = V = \text{Linear}(u(i,j)_k^{(t-1)})
\end{equation}
where the same multi-head self-attention and FFN mechanisms are applied to the edge features,
\begin{equation}
u(i,j)_k^{(t)} = \text{FFN}(u(i,j)_k^{(t)})
\end{equation}

After updating the node and edge features with the intra-crystal and inter-atomic dual-Transformer layers, the model concatenates the features by,
\begin{equation}
z_{i,j,k}^{(t)} = v_i^{(t)} \oplus v_j^{(t)} \oplus u(i,j)_k
\end{equation}
where \(\oplus\) means concatenation and \(z_{i,j,k}^{(t)}\) represents the combined feature of atoms and edges.

Then the atom feature vectors are updated through a non-linear graph convolution function,
\begin{equation}
v_i^{(t+1)} = v_i^{(t)} + \sum_{j,k}\sigma(z_{i,j,k}^{(t)} W_f^{(t)} + b_f^{(t)}) \odot g(z_{i,j,k}^{(t)} W_s^{(t)} + b_s^{(t)})
\end{equation}
where \( g \) denotes the activation function, \( \sigma \) is the sigmoid function served as a gate, \( \odot \) denotes element-wise multiplication, \( W_f^{(t)} \) and \( W_s^{(t)} \) are the convolution weight matrices, and \( b_f^{(t)} \) and \( b_s^{(t)} \) are the bias vectors.

After \( R \) convolutional iterations, the network updates the feature vector \( v_i^{(R)} \) for each atom. A pooling layer then aggregates these vectors into a global feature vector \( v_c \) for the entire crystal by,
\begin{equation}
v_c = \text{Pooling}(\{v_i^{(R)} | \ i \in \text{atoms}\})
\end{equation}
The pooled vector \( v_c \) is a vector containing all the information of the subgraphs of the crystal. It is permutation invariant, so it can capture the information ignoring the noise and rotate translation. It is then passed through fully-connected layers to predict the target property \( \hat{y} \), with the training process minimizing the cost function \( J(y, \hat{y}) \), representing the difference between the predicted property \( \hat{y} \) and the DFT-calculated property \( y \).

\subsection{Crystal Angular Encoder}

To enrich the information of edges, we go beyond merely distance information by adding angular information into the edges, which allows for a more detailed depiction of the spatial relationships between atoms. For a given atom $i$ and its neighbor $j$, we calculated the related angles $\theta_x$, $\theta_y$, and $\theta_z$  between the edge vector $\mathbf{r}_{ij}$ and the axes vectors $\mathbf{x}$, $\mathbf{y}$, and $\mathbf{z}$ by,
\begin{equation}
\theta_x = \cos^{-1}\left(\frac{\mathbf{r}_{ij} \cdot \mathbf{x}}{\|\mathbf{r}_{ij}\| \|\mathbf{x}\|}\right),
\end{equation}

\begin{equation}
\theta_y = \cos^{-1}\left(\frac{\mathbf{r}_{ij} \cdot \mathbf{y}}{\|\mathbf{r}_{ij}\| \|\mathbf{y}\|}\right),
\end{equation}

\begin{equation}
\theta_z = \cos^{-1}\left(\frac{\mathbf{r}_{ij} \cdot \mathbf{z}}{\|\mathbf{r}_{ij}\| \|\mathbf{z}\|}\right).
\end{equation}

These angles are then encoded into a feature vector using an angular encoder, which discretizes each angle into one of several predefined bins, written as, 

\begin{equation}
\text{Encoded}_{\theta_x}[k] = 
\begin{cases}
1, & \text{if }\ \  k\cdot\Delta\theta \leq \theta_x < (k+1)\cdot\Delta\theta \\
0, & \text{otherwise}
\end{cases},
\end{equation}

where $\displaystyle \Delta\theta = \frac{2\pi}{\text{bins}}$ and $k$ ranges from 0 to $\text{bins}-1$. Analogous encoding processes apply to $\theta_y$ and $\theta_z$.
The edge feature vector which combining RBF (Radial Basis Function) and angular features, is used as the edge feature which is further processed in the model. 

\section{benchmark}

In order to evaluate the performance, we used JARVIS-DFT\cite{choudhary2014jarvis}, dated 2021.8.18 as the training database. The dataset comprises 25,922 materials with bandgap, formation energy, and etc. For training, validation and testing splits, JARVIS-DFT\cite{choudhary2014jarvis} database and its properties are split into 60\% training, 20\% validation, and 20\% testing sets. To further evaluate the performance, we merge two distinct datasets of perovskite materials \cite{kim2017hybrid,nakajima2017discovery} to create a more diverse and representative dataset which contains 3489 perovskite structures with formation energy and bandgap, key properties for perovskites. For comparison, the state-of-the-art GNN models of CGCNN and MEGNET are also used to model the formation energies and bandgaps of the materials involved in the teo materials database, and the resuts are listed in Table\ref{tab:bench}

\begin{figure}[H]
    \centering
    \includegraphics[width=1\linewidth]{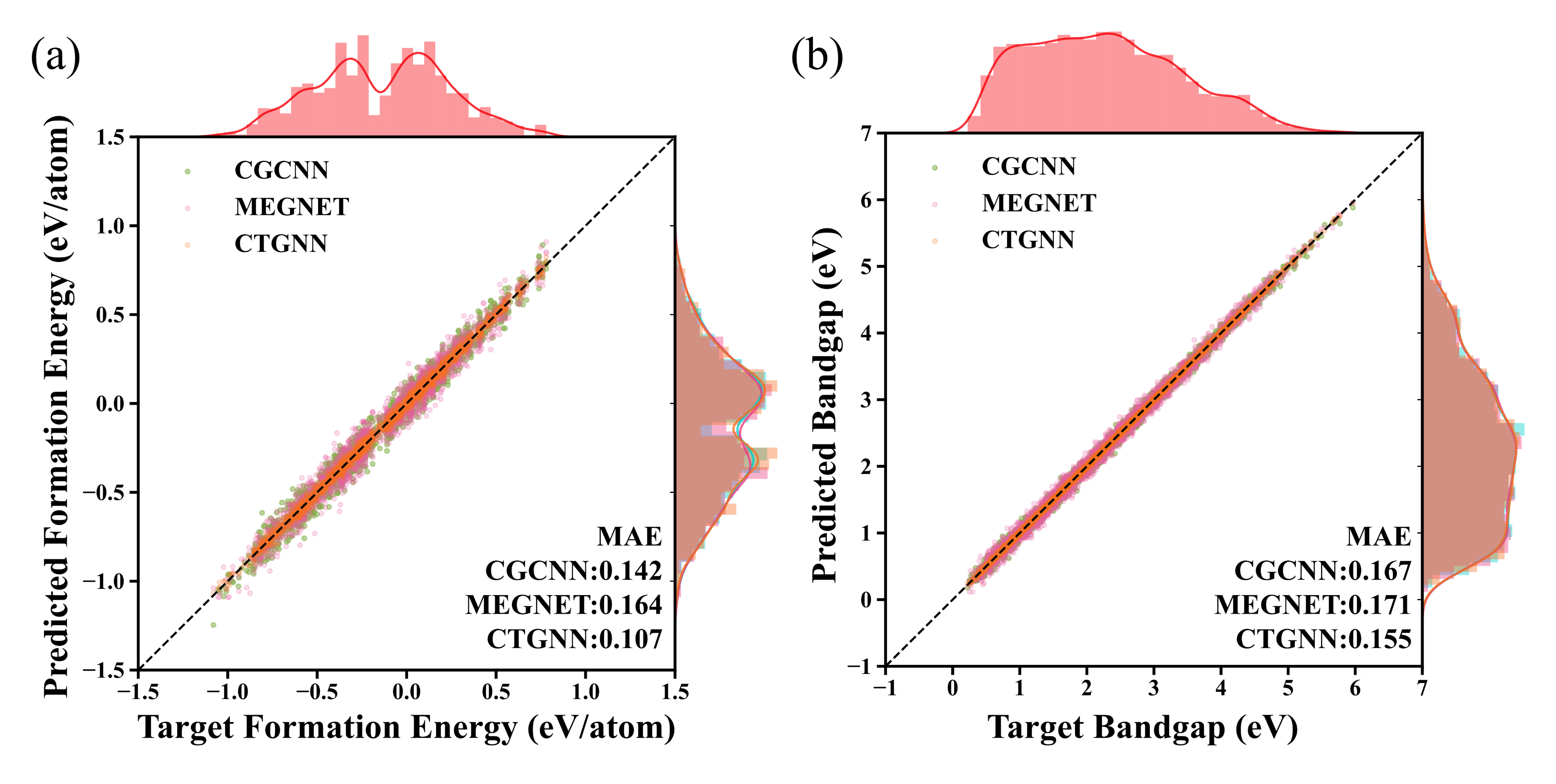}
    \caption{Plots of predicted formation energy and bandgap versus target for CGCNN, MEGNet, and CTGNN models, respectively. the upper and right part are the target and prediction data distribution.}
    \label{fig:scatter}
\end{figure}

\begin{table}[H]
\centering
\caption{benchmark on jarvist dataset and perovskite dataset. $E_f$ and $E_g$ denotes formation energy and bandgap respectively.}
\label{tab:bench}
\begin{tabular}{p{5cm}p{3cm}p{3cm}p{3cm}}
\hline
\textbf{target} & \textbf{model} & \textbf{MAE} & \textbf{R$^2$} \\ \hline
\multirow{3}{*}{Pero($E_f$) (eV/atom)} & CGCNN & 0.027 & 0.988 \\
 & MEGNet & 0.032 & 0.982 \\
 & CTGNN & 0.013 & 0.996 \\ \hline
\multirow{3}{*}{Pero($E_g$) (eV)} & CGCNN & 0.285 & 0.855 \\
 & MEGNet & 0.296 & 0.845 \\
 & CTGNN & 0.156 & 0.960 \\ \hline
\multirow{3}{*}{Jarvist($E_g$) (eV)} & CGCNN & 0.531 & 0.914 \\
 & MEGNet & 0.493 & 0.908 \\
 & CTGNN & 0.469 & 0.910 \\ \hline
\end{tabular}
\end{table}
As listed in Table \ref{tab:bench}, our proposed CTGNN model demonstrates superior performance on the perovskites dataset on formation energy and bandgap prediction. The plots of the target and prediction distribution are also shown in Fig \ref{fig:scatter}. Specifically, CTGNN achieves the lowest MAE on the formation energy prediction on the Pero dataset, compared to CGCNN and MEGNet models. With the MAE of 0.013 eV/atom, CTGNN is 51.85\% and 59.38\% better than CGCNN and MEGNet, whose MAE is 0.027 and 0.032 eV/atom  respectively. The R$^2$ is also the highest, with 0.8\% and 1.4\% improvements. When it comes to the bandgap prediction on the Pero dataset, CTGNN also surpasses CGCNN and MEGNet models with a lower MAE and higher R$^2$. To be more detailed, the MAE is 0.156 eV, which is 45.26\% and 47.30\% better. The Jarvist dataset exhibits the same trend, with the CTGNN model enjoying the lowest MAE of 0.469 eV, which is 11.67\% and 4.87\% better, a significant improvement. These results underline the effectiveness of CTGNN in handling complex material datasets, especially in the perovskites datasets over traditional methods like CGCNN and MEGNet, highlighting its potential in the perovskites era.

\section{Ablation Study}

To understand the contribution of different components in the CTGNN model, we conducted an ablation study on the Pero dataset for bandgap prediction. The study involved removing key components from the model and evaluating the resulting performance. The results are summarized in Table \ref{tab:ablation}.

\begin{table}[H]
\centering
\caption{Ablation study on Pero dataset for bandgap prediction.}
\label{tab:ablation}
\begin{tabular}{p{8cm}p{3cm}p{3cm}}
\hline
\textbf{Model} & \textbf{MAE} (eV) & \textbf{R$^2$} \\ \hline
CTGNN (full model) & 0.156 & 0.960 \\ 
Without Angular Encoding & 0.188 & 0.946 \\ 
Without inter-atomic Transformer & 0.190 & 0.945 \\ 
Without Transformer & 0.285 & 0.855 \\ \hline
\end{tabular}
\end{table}

As shown in Table \ref{tab:ablation}, removing the angular encoding and the neighbor Transformer from the CTGNN model leads to a decrease in performance. Specifically, without the angular encoding, the MAE increases to 0.188 eV and the R$^2$ decreases to 0.946. Similarly, without the inter-atomic Transformer, the MAE increases to 0.190 eV and the R$^2$ decreases to 0.945. When the dual-Transformer is totally removed The performance drop to 0.285 from 0.190.

\section{Conclusion}
CTGNN model represents a significant progress in the field of material computing, particularly in perovskite materials. By innovatively combining the advantages of Transformer model and graph neural networks, CTGNN can capture both the local and global interactions in materials efficiently. The addition of angular kernels allows for a more comprehensive representation of atomic structures, surpassing the traditional models which only focus on distances. Our results demonstrate that CTGNN outperforms existing models in predicting key material properties such as formation energy and bandgap, which is confirmed by the benchmark tests on multiple datasets. CTGNN not only enhance the ability to predict material properties with greater accuracy, but also provide a solid foundation for discovering new materials.

\hspace{1 mm}

\section*{Acknowledgements}
This work is supported by the National Key R\&D Program of China (2023YFA1608501), and Natural Science Foundation of Shandong Province
under grants no. ZR2021MA041. Mr. L. Jin and Z. Du also want to acknowledge the support of FDUROP (Fudan's Undergraduate Research Opportunities Program) (24052, 23908).

\subsection*{Competing Interests statement}

The authors declare no competing interests.

\subsection*{Author Contributions}


\end{document}